\documentclass[aps,pra,twocolumn,superscriptaddress,groupedaddress]{revtex4-1} 
\usepackage{graphicx}  
\usepackage{dcolumn}   
\usepackage{bm}        
\usepackage{amssymb}   
\usepackage{amsmath}
\usepackage{transparent}
\usepackage{hyperref}

\begin{document}

\title{Spectroscopic detection of atom-surface interactions in an atomic vapour layer with nanoscale thickness}

\author{K. A. Whittaker$^1$} 
\author{J. Keaveney$^1$}
\author{I. G. Hughes$^1$}
\author{A. Sargsyan$^2$}
\author{D. Sarkisyan$^2$}
\author{C. S. Adams$^1$}
\address{$^1$Joint Quantum Centre (JQC) Durham-Newcastle, Department of Physics, Durham University,South Road, Durham, DH1 3LE, United Kingdom
\\$^2$Institute for Physical Research, National Academy of Sciences, Ashtarak 2, 0203, Armenia}

\begin{abstract}

We measure the resonance line shape of atomic vapor layers with nanoscale thickness confined between two sapphire windows. The measurement is performed by scanning a probe laser through resonance and collecting the scattered light. The line shape is dominated by the effects of Dicke narrowing, self-broadening, and atom-surface interactions. By fitting the measured line shape to a simple model we discuss the possibility to extract information about the atom-surface interaction.

\end{abstract}

\maketitle

\section{Introduction}

Atomic alkali vapor cells are widely used in applications ranging from magnetometry~\cite{Griffith_magnetom}, sources of quantum light~\cite{KrupkeAlkaliLaser}, electric field imaging~\cite{Fan2014}, atomic clocks~\cite{Knappe2004}, nanophotonics~\cite{Stern2013}, optical isolators~\cite{Weller2012b} and quantum memory~\cite{Lvovsky09}. Modern vapor cell technology has been trending towards integrating miniaturized cells into the aforementioned applications, with millimeter~\cite{LiewNIST,StraessleInCell,Woetzel2013,FemtoSt}, micron~\cite{Baluktsian_VCfab} and even nanometer scale cells~\cite{Sarkisyan2001ETC} popular areas of investigation. An accurate method of analyzing spectra from miniature cells is vital to aid the development of such technology.

If the vapor is confined spatially with a dimension less than the transition wavelength fascinating new physical effects become accessible such as the narrowing of spectral lines~\cite{Dicke1953}, extreme dispersion resulting in large negative group indices and superluminal propagation~\cite{Keaveney2012a,Whittaker2015}, repulsive van der Waals interactions~\cite{Failache1999}, the cooperative Lamb shift~\cite{Keaveney2012CLS}, and perhaps a medium where the Kramers-Kronig relations are violated~\cite{Wang_KKviol}.

The tight confinement of the atoms inside nanocells opens up opportunities to study the interaction of atoms with a nearby surface, explored using spectroscopy on both low-lying~\cite{Whittaker2014} and higher-lying excited states~\cite{Hamdi2005} or EIT spectroscopy of highly-excited Rydberg states~\cite{Kubler2013}. This can be expanded to investigate the temperature dependence of the coefficients describing the strength of the atom-surface (AS) interaction~\cite{Bloch2014_Tdep} and cases where the usually attractive AS interaction becomes repulsive due to surface resonances~\cite{Failache1999}. Alternative methods to spectroscopy for probing the AS interaction have also been used, such as scattering or deflection of a beam close to a surface~\cite{Bender2010,Pasquini2004,Druzhinina2003,Shimizu2001,Sukenik1993patch_CP,Sandoghdar1992}, atomic beam diffraction~\cite{Grisenti1999,Perreault2005} and reflection of an ultracold atom cloud from an atomic mirror~\cite{Landragin1996,Mohapatra2006b}. Such experiments use detection methods that take place some time after the interaction has occurred. 

In this paper we employ spectroscopic measurements that allow us to infer the effects of the AS interaction at the time of absorption. Note that in this case, both ground and excited states contribute to the measurement of the AS interaction. In what follows we present detailed spectra for a range of  cell lengths and temperatures, expanding upon a previous publication~\cite{Whittaker2014}, giving further detail into the error analysis and fitting procedures, present a simulation justifying the model used to model velocity selective effects used and offer further analysis on the results found. We describe an experiment that takes many absorption spectra over a range of temperatures and cell lengths, and uses fitting and error analysis to precisely measure the general form of the AS interaction within the near field. We first detail the methods of data acquisition for absorption spectra, where we utilize single photon counting modules (SPCMs) to acquire absorption spectra over long integration times. The ease of acquisition allows us to take many spectra for a range of cell lengths and temperatures. We then discuss how experimental error is accounted for and utilized to increase the precision of our measurements. 

The theory section details the form of the AS interaction in the near field (van der Waals) regime~\cite{LennardJonesAS}, discussing possible effects that may alter the final form, such as a transition between the near field and retarded (Casimir Polder)~\cite{Casimir1948} regimes; the effect of multiple reflections and the effect of temperature on the interaction. We then give an overview of the fitting procedure performed on the spectra taken. The model we have developed accounts for self broadening~\cite{Weller2011a}, Dicke narrowing effects~\cite{Dicke1953} and the AS interaction~\cite{LennardJonesAS}. The results section presents fitting results for spectra taken on the Rb D2 line and the Cs D1 line, where we find that to describe spectra in the length range investigated herein, the van der Waals description of the AS interaction is optimal. We finally discuss the implication and accuracy of our results including a discussion of the limitations of the technique and perspectives for further work.

\section{\label{sec:exp_err}Experiment }

\subsection{Data Acquisition}

\begin{figure}
\includegraphics[scale = 0.3]{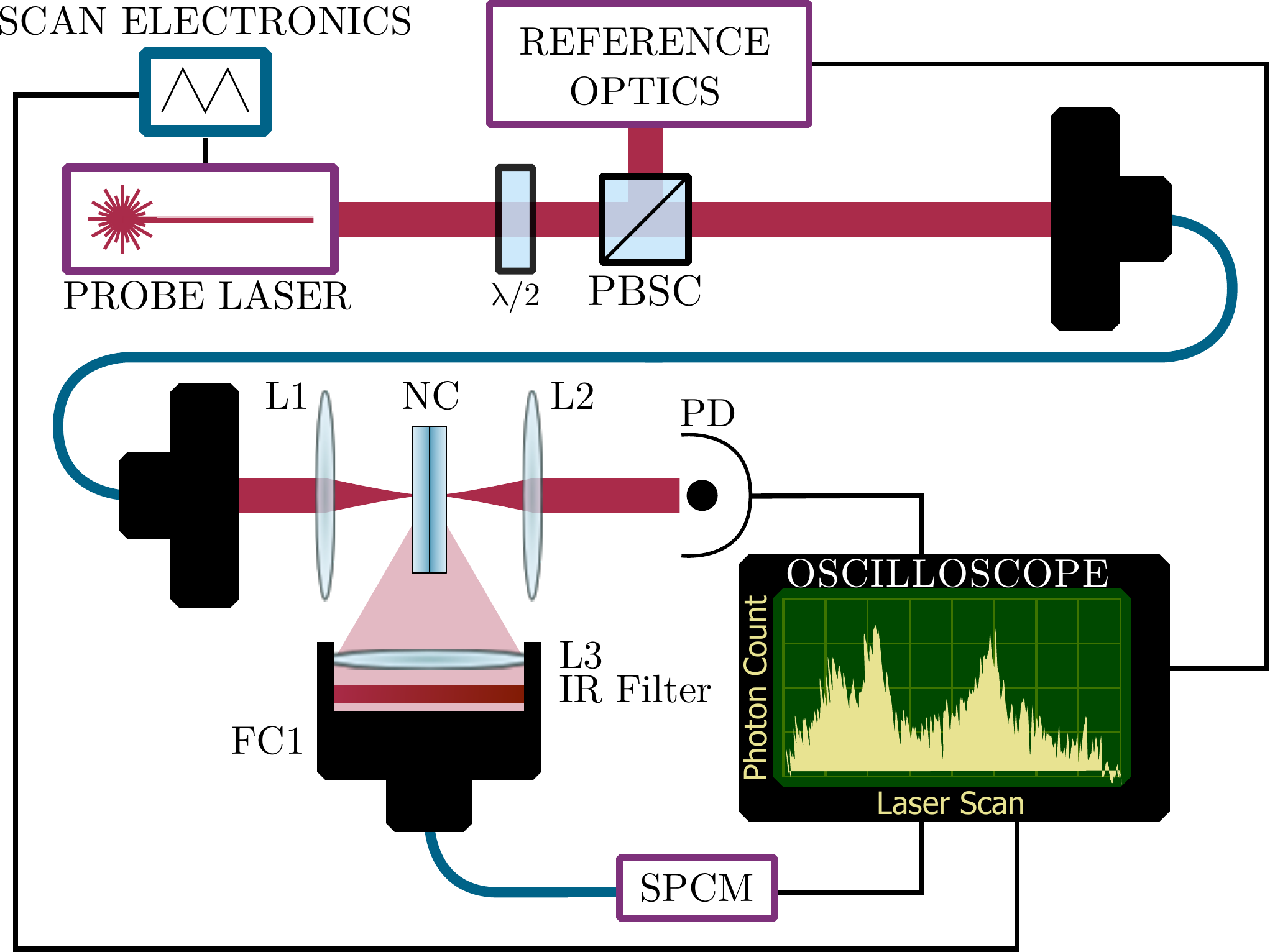}
\caption{\label{fig:exp_setup}Experimental schematic used to measure the atomic line shape. The laser is scanned through resonance and we detect the scattered light. Part of the beam is split off by a half-wave plate ($\lambda/2$) and polarizing beam splitter cube (PBSC) to the reference optics for frequency calibration. The beam is then fibered to the nanocell (NC) section of the experiment, where lens L1 focuses the beam to a $1/\mathrm{e}^2$ radius of 20 $\mu$m inside the NC. The transmitted beam is re-collimated by lens L2 and a transmission spectra is collected on photo-diode PD. Off-axis scattered photons are collimated by lens L3, with ambient light filtered out by an infrared (IR) filter, and sent into a single photon counting module (SPCM) via an fiber collimator FC1. The signal is processed by the oscilloscope to generate a histogram of the photon arrival times.} 
\end{figure}
\begin{figure}
\includegraphics[scale = 0.42]{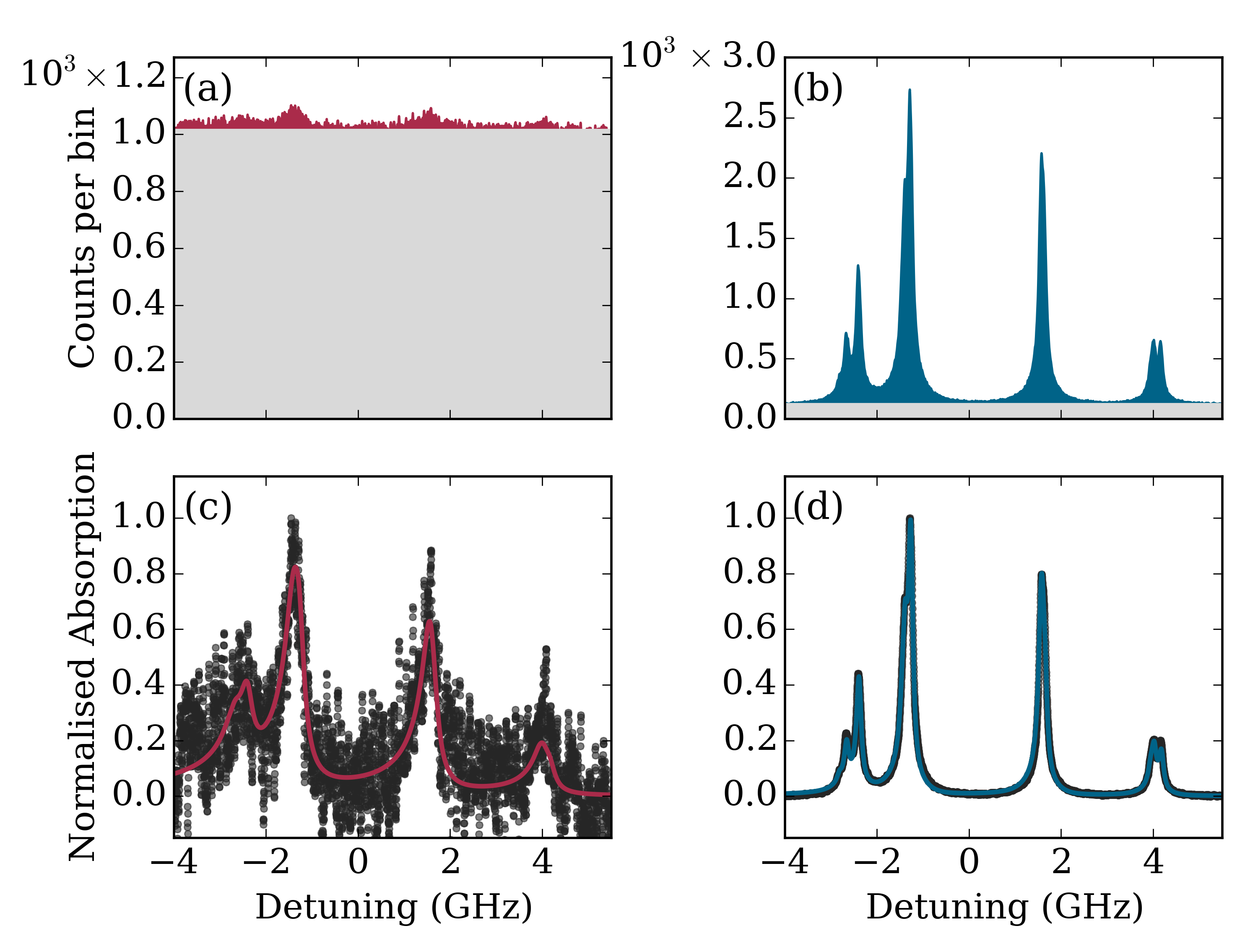}

\caption{\label{fig:raw_rb}Off-axis scattered photons measured using an excitation laser scanned across the Rb D2 resonance. The figure shows a comparison of raw data (gray areas, panels (a) and (b)) to processed data (black points, panels (c) and (d)) on the Rb D2 line at a temperature of 125$^\circ$C for (a,c) 60~nm and (b,d) 250~nm cell lengths. Background counts are shown in gray and the atomic signal highlighted in red and blue for lengths of 60 and 250 nm respectively. Panel (a): The atomic signal, highlighted in red, is small in comparison to the background count because the interrogation volume is so small that there are relatively few atoms contributing to the signal. Therefore, the error is dominated by the error of the background count. Panel (b): The atomic signal is the main component of the raw signal because far more atoms are contributing to the signal, resulting in much smaller relative error compared to panel (a). Panel (c): The fit for $L$~=~60~nm (red line) has a `goodness of fit' parameter, $\Upsilon$, of 0.1. The small value of $\Upsilon$ is caused by the large background counts involved for such small cell lengths, illustrated in panel (a). Panel (b): the fit for $L$~=~250~nm (blue line) has a reduced chi squared $\Upsilon$~\cite{hughes2010measurements} of 1.7, indicating an excellent fit. The background photon count is much smaller than for $L$~=~60~nm, resulting in lower relative error and an $\Upsilon$ closer to 1.}
\end{figure}

To probe the atom surface (AS) interaction we use nanocells (NCs), produced at the National Academy of Science in Armenia~\cite{Sarkisyan2001ETC}. The cells are made from c-axis aligned sapphire with an interior surface roughness of 3~nm and are described in more detail in previous publications~\cite{Keaveney2012a,Keaveney2012CLS,KeaveneyThesis,opacity_sat}.  Each cell has a wedged profile, with a thickness range of 2~$\mu$m to 30~nm. In order to exploit variable vapour column thickness in the range of 30-2000~nm, the cell is vertically wedged by placing a 2~$\mu$m thick platinum spacer strip between the windows at the bottom side prior to gluing. 

The NC thickness, $L$, is measured using back reflections from the surfaces inside the cell. Two back reflections R1, from the front surface of the cell, and R2, a combination of reflections from both interior surfaces, are measured. Etalon effects inside the cell mean the two beams that compose R2 interfere constructively or destructively depending on the cell length, resulting in a varying intensity of R2. The thickness is calculated by taking the ratio of intensities R1 and R2 and applying a standard Fabry-Perot treatment~\cite{silfvast2004laser}.

Our method measures absorption or more precisely extinction in the forward scattering direction by detecting the off-axis scattered light resulting in absorption spectra that we can fit to measure the AS interaction. We use this method because a measurable transmission profile for extremely short cell lengths can only be obtained by heating to high temperatures (exceeding 300$^\circ$C), where collisional broadening smears out potential atom surface measurements \cite{Weller2011a}. For shorter cell lengths (L $<$ 100~nm) and lower temperatures (T $<$ 200$^\circ$C), absorption is less than 0.5\% of the incident light, making the measurement of transmission impractical. Using off-axis detection of scattered light with single-photon counters, extremely high sensitivity spectra can be taken from shorter cell lengths at lower temperatures. This increases our experimental resolution as spectra with all hyperfine states resolved can be taken using single-beam spectroscopy. The narrowing of the lines is caused by Dicke narrowing~\cite{Dicke1953} effects inside the cell- a motional effect that will be outlined in section \ref{sec:theory_fit}. Absorption spectra were taken on the Rb D2 line; used for its large AS interaction on the ground state transition, and the Cs D1 line; used for its large 9 GHz ground state splitting and 1.2 GHz excited state splitting, much larger than the Doppler width. All spectra were taken within the weak-probe limit~\cite{Sherlock_weakprobe}.

The experimental layout is shown in Fig.~\ref{fig:exp_setup}. Some of the light is sent though reference optics - a 7.5~cm reference vapour cell and a Fabry-Perot etalon, which are collected concurrently with the scattered light spectra to calibrate the laser frequency. The light is then sent down an optical fiber to the NC section of the experiment. The light is focussed in the centre of the NC by lens L1 to a spot size with a $1/\mathrm{e}^2$ radius of 20~$\mu$m, then is recollimated by lens L2. Transmission data are recorded on a photodiode (PD). Off-axis scattered light is collected by lens L3, with background light and thermal photons from the cell heater filtered out by an infrared filter. The signal is sent to a single-photon counting module (SPCM) via a fiber collimator, FC3. The counts are processed by a LeCroy Waverunner 610Zi oscilloscope into a histogram of arrival times corresponding to the frequency of the laser scan, according to the method outlined in~\cite{KeaveneyThesis}. 

The laser scan time is 50~ms, and individual bins in the histogram are 5~$\mu$s long, corresponding to a bin width of 2~MHz. Hence the detection method is sensitive to the frequency at which a photon was absorbed, meaning AS shifts measured map directly onto a particular AS separation for that frequency. Our method is insensitive to any motion between the time of absorption and fluorescence events. Single scans return only a few photon counts, so long integration times (15~minutes to several hours) are used to build up statistics of arrival times, allowing for sensitive detection of spectra.

Figs. \ref{fig:raw_rb}(a) and (b) show examples of raw spectra taken for $T$ = 125$^\circ$C. The data are taken at 2 different cell lengths, 60~nm (red area) and 250~nm (blue area). The raw spectra have a very large background count, highlighted in grey. Signal-to-background ratio varies hugely between different cell lengths, only the atomic signal is highlighted (colored areas) in Figs.~\ref{fig:raw_rb}(a) and (b). The effect of these background counts on experimental errors will be discussed in the next subsection. Before analysis, the background signal is subtracted from the atomic signal. The resulting processed data is shown as black points in Figs. \ref{fig:raw_rb}(c) and (d). We fit the spectra, using a procedure outlined in section~\ref{sec:theory_fit}. Examples of fit results are shown on Figs.~\ref{fig:raw_rb}(c) ($L$~=~60~nm, red line) and (d) ($L$~=~250~nm, blue line).

\subsection{Error analysis}

The errors in the atomic signal, $A(\omega)$, are calculated by taking the frequency-dependent raw spectrum $R(\omega)$ and subtracting the background count $B$.
\begin{equation}
A(\omega) = R(\omega)-B.
\end{equation}
For data acquisition, the errors in the atomic signal ($s_{\mathrm{A}}$) are Poissionian counting statistics; each data point in the raw counts $R(\omega)$, with a number of counts $N_\mathrm{count}$, has an associated error of $\sqrt{N_\mathrm{count}}$~\cite{hughes2010measurements}. The errors can be calculated as:
\begin{equation}
s_\mathrm{A} = \sqrt{s_\mathrm{R}^2+s_\mathrm{B}^2},
\end{equation}
with $s_\mathrm{R}$ and $s_\mathrm{B}$ as the errors in the raw and background signals, respectively. The relative error bars are very small for spectra similar to Fig.~\ref{fig:raw_rb}b because $A(\omega)$ is large and $B$ is small. However for spectra with low signal-to-background ratio like that in Fig.~\ref{fig:raw_rb}(a), i.e. $B \gg A(\omega)$, the error of the raw signal is roughly equal to the error of the background signal, $s_\mathrm{R}\approx s_\mathrm{B}$. Hence, $s_\mathrm{A}$ can be approximated as $s_\mathrm{A} \approx \sqrt{2s_\mathrm{B}^2 }$. This results in large relative error for shorter cell lengths when compared to the error bars for longer cells with better signal-to-background ratios.

We fit the spectra using the Marquardt-Levenburg method~\cite{hughes2010measurements} to minimise the sum of the squares of the difference between theory and experiment, normalized by the error bar. For errors in fitted parameters for an individual spectrum, we use the standard deviations of the mean calculated by the fitting routine.

For the analysis performed on Rb spectra, we take an average of the fitted parameters weighted by the reduced $\chi$-squared of the fit. The reduced $\chi$-squared is a goodness of fit measurement, represented here with $\Upsilon$ to avoid confusion with the vapor susceptibility. $\Upsilon$ is calculated as
\begin{equation}
\Upsilon = \frac{1}{\nu} \sum_\omega \frac {[A(\omega)-T(\omega)]^2}{s_A^2},
\end{equation}
where $A(\omega)$ is the observed frequency-dependent atomic signal, $T(\omega)$ is the theoretical signal and $\nu$ is the degrees of freedom, calculated as $\nu = N_\mathrm{points} - N_\mathrm{param}$, where $N_\mathrm{points}$ is the length of the data set and $N_\mathrm{param}$ is the number of fitted parameters.

Values for $\Upsilon$ of our fitted spectra vary between 0.1 to over 1000. Generally, a $\Upsilon$ close to 1 indicates a good fit, and any value much lower than 1 indicates overestimated errors~\cite{hughes2010measurements}. However, this is not the case in our experiment; the smaller values of $\Upsilon$ are caused by the huge background counts for shorter cell lengths, illustrated in Figs. \ref{fig:raw_rb}(a) and (b). The small values of $\Upsilon$ can be accounted for by considering the large relative error margins that arise when the atomic signal is small resulting in a higher proportion of the fitted spectra within the error margins of the atomic signal. Consequently, the resulting $\Upsilon$ is much less than 1, but still a valid measure of goodness of fit.

\section{\label{sec:theory_fit}Theory and Fitting}

\subsection{The Atom-Surface Interaction}

\begin{figure}
\includegraphics[scale = 0.5]{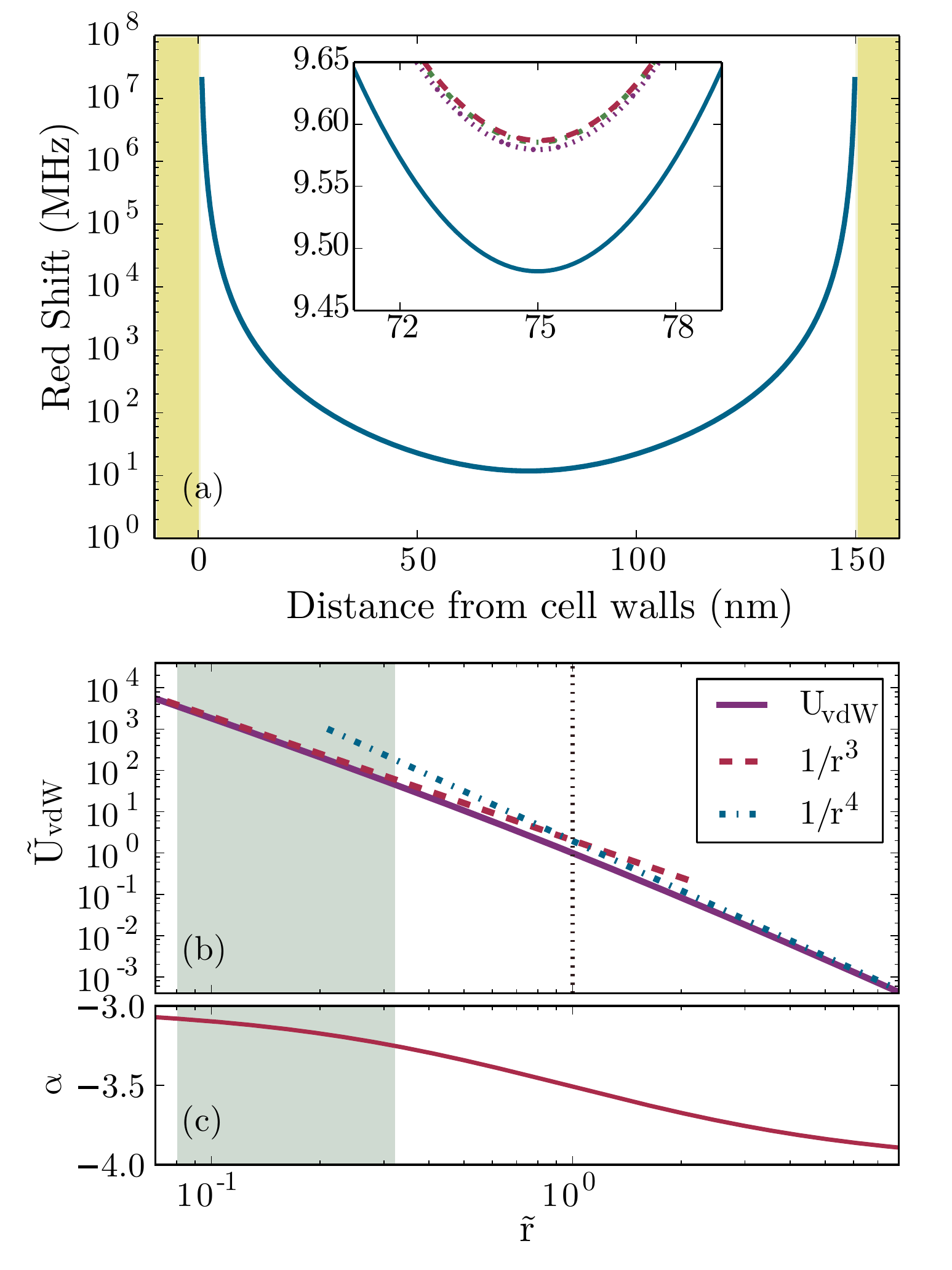}
\caption{\label{fig:AS_diag}(a) The position-dependent red shift from the vdW $1/r^3$ AS potential. Inset: The effect of multiple reflections contributing to the vdW AS interaction. Plotted are shifts caused by accounting for a single (blue line); two (purple dotted line); three (red dashed) and four reflections (green dot dashed line). These differences are on the order of 100 kHz, smaller than the range of shifts our experiment is sensitive to. Panel (b) shows the length-dependent variation of AS potential (purple line) taken from equation (45) in~\cite{Friedrich_1992}, highlighting the transition between the short range van der Waals and long range Casimir-Polder regimes. At short lengths, there is good agreement with a $1/r^3$ vdW potential (red dashed line). At longer lengths there is reasonable agreement with a $1/r^4$ CP potential (blue dot dashed line). $\tilde{U}_\mathrm{vdW}$ is scaled such that $\tilde{U}_\mathrm{vdW} = 1$ at $\tilde{r} = \lambda/2\pi$, and $\tilde{r}$ is the AS distance in units of $\lambda/2\pi$. Most of our data come from the region highlighted in gray (length range of 10-40 nm). In panel (c), the variation of the exponent $\alpha$ of the atom-surface attraction in the form $1/r^\alpha$ between the vdW and CP regime is shown, again with the length range the experiment is sensitive to highlighted in gray.}
\end{figure}

The atom-surface (AS) interaction is the attraction between a dipole and its reflection in a surface. An induced dipole in the atomic medium induces a shift of charge in a nearby conducting or dielectric surface~\cite{LennardJonesAS}. This charge redistribution results in a virtual dipole image~\cite{Casimir1948}, which has an attractive interaction with the original atomic dipole. This causes a red-shift of the atomic transitions, generally described using $-C_\alpha/r^\alpha$, where $\alpha$ is an exponent that varies between 3 and 4, depending on the interaction regime. 

The Casimir-Polder (CP) interaction describes this process at AS separations $r > \lambda/2\pi$, where $\lambda$ is the transition wavelength. The interaction between the real and image dipole is retarded by the transit time of photons~\cite{LennardJonesAS}, and the potential takes the form $U_\mathrm{CP} = -C_4/r^4$, with $C_4$ as a coupling coefficient determining the strength of the interaction. 

When the AS separation reaches the regime $r < \lambda/2\pi$, the induced shifts follow the van der Waals description $U_\mathrm{vdW} = -C_3/r^3$, where $C_3$ is the coupling coefficient of the van der Waals interaction. Fig.~\ref{fig:AS_diag}(a) shows the induced red shift caused by $U_\mathrm{vdW}$.

$C_3$ is dependent on the atomic polarizability~\cite{Friedrich_1992}, and hence varies depending on the electronic transition the shift is acting upon. When measuring the AS shift on an optical transition, the effective coupling coefficient is the difference between the ground and excited states. For this reason, we distinguish between coupling coefficients using a superscript denoting the transition, e.g. $C_3^{5S_{1/2}}$ for the $^{85}$Rb 5$S_{1/2}$ $F_g$ = 2,3 ground states and $C_3^{5S_{1/2}\rightarrow5P_{3/2}}$ for measurements taken on the $^{85}$Rb 5$S_{1/2}$ $F_g$ = 2,3 to $^{85}$Rb 5$P_{3/2}$ $F_e$ = 1,2,3,4 transition.

The smooth transition between the vdW (near field) and CP regime~\cite{Caride2005} has been the subject of many studies~\cite{Casimir1948,Friedrich_1992,Eizner_CPvdW}, with estimates for the onset of the transition ranging from lengths of $\lambda/2\pi$~\cite{Friedrich_1992} to 0.03$\lambda$~\cite{desorption}. To identify the length regime our work covers, we inspect the full form of the AS interaction including both CP and vdW forms~\cite{Friedrich_1992}:
\begin{equation}
U_\mathrm{AS} = -\frac{\hbar^2}{2M}\left[\frac{r^3}{\beta_3}+\frac{r^4}{\beta_4}\right]^{-1},
\end{equation}
where $M$ is the atomic mass and $\beta_{3,4}$ are length parameters related to the strength of the vdW/CP potential and are taken from~\cite{Friedrich_1992}. Values taken are for ground state Rb (5$S_{1/2}$), unlike the transitions we investigate experimentally, which probe the difference in interaction coefficients between the ground (5$S_{1/2}$) and excited (5$P_{3/2}$) states. Calculation of $\beta_{3,4}$ for the excited state transitions are beyond the scope of this study, and we use the information from Fig.~\ref{fig:AS_diag}(b) as an indicator for which interaction regime we expect our spectra to lie in. We expect that the ground-state case is a worse case scenario as we are further into the $C_3$ region for the excited state as the transition wavelengths to nearby states are longer. Fig.~\ref{fig:AS_diag}(b) shows the resulting potential for the ground state (purple line). The length scale $\tilde{r}$ is scaled in units of $\lambda/2\pi$. $U_\mathrm{vdW}$ is normalized such that $U_\mathrm{vdW} = 1$ at $\tilde{r} = 1$. Both the CP and vdW interactions are plotted for comparison (blue dot-dashed and red dotted lines respectively). The length range that most of our data are produced for is shown as a gray area on the plot. In Fig.~\ref{fig:AS_diag}(c) we plot the length dependence of the exponent $\alpha$ of the AS interaction. Figs. \ref{fig:AS_diag}(b) and (c) demonstrate that our experiment is sensitive to the shorter range van der Waals interaction regime. The average exponent in the length range $L=10-40$~nm is $\alpha=3.16$. 

The impact of the AS interaction on the final resonance line shape is usually an overall red-shift of the line shape, accompanied by a long asymmetric red tail due to atoms at different positions inside the cell experiencing a different shift. However, if surface polariton resonances coincide with the atomic resonance, the AS interaction becomes repulsive, instead causing a blue shift of the atomic resonance~\cite{Failache1999}.

The values of $C_3$ measured in this experiment are for the Rb $5S_{1/2} \rightarrow 5P_{3/2}$ transition. The $C_3$ coefficient for the transition when the atoms are interacting with a perfectly conducting surface is $C_3^{5S_{1/2}\rightarrow5P_{3/2}} = 4.1 \pm 0.2$ kHz $\mu$m$^{3}$~\cite{priv_comm}, calculated using the methods outlined in~\cite{Scheel2008}. When the surface is not a perfect conductor, $C_3$ is multiplied by a reflection coefficient $R$~\cite{Fichet1995}. Sapphire has no significant variation in the frequency-dependent permittivity $\epsilon(\omega)$ in the relevant frequency range, and hence we use a value of $\epsilon_\mathrm{sapph}(\omega) = 3.24 $ for the Cs D1 and Rb D2 line, hence the reflection coefficient can be expressed as $R = (\epsilon_\mathrm{sapph}(\omega)-1)/(\epsilon_\mathrm{sapph}(\omega)+1)$. This results in a scaling factor of 0.53 for values of $C_3^{5S_{1/2}\rightarrow5P_{3/2}}$ measured in this experiment.

The coupling coefficient $C_3$ also varies with temperature~\cite{Caride2005,Bloch2014_Tdep}. However at short ranges the thermal dependence of $C_3$ is dominated by surface excitations~\cite{Bloch2014_Tdep}. This is not the case for the transitions investigated herein, as they do not concur with the frequencies of any surface plasmon excitations of sapphire~\cite{Fichet1995}. There is still some temperature variation of the $C_3$ coefficient, this is on the order of 1\% over the temperature ranges investigated herein and is hence negligible in the scope of this experiment~\cite{priv_comm}.

The AS interaction is also affected by multiple reflections in the surface, where the induced dipole interacts with reflections at integer multiples of the AS separation $r$. The effect of such reflections are shown in the inset to Fig.~\ref{fig:AS_diag}a. The potential from a single reflection (blue line) differs only slightly from the effect of two (purple dotted line), three (red dashed) and four (green dot-dashed) reflections. These differences are on the order of a few hundred kHz, smaller than detectable in the experiment.

\subsection{Line shapes for absorption spectra in nanometric cells}

\begin{figure*}
\includegraphics[scale = 0.6]{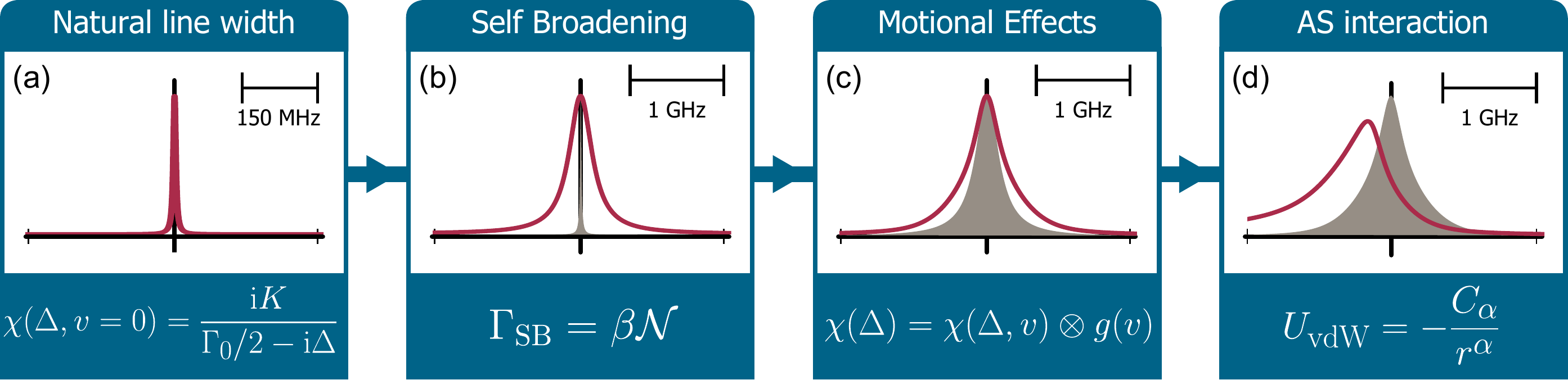}
\caption{\label{fig:fitting_process}Illustration of the various contributing factors to the measured line-shapes. From left to right, the line-shape after each effect (red line) is added to the previous line shape (shown as grey area). Each step is accompanied by the most relevant equation used to describe the changes in the line shape. The spectra are generated for a cell length of 60 nm and temperature of 200$^\circ$C. Panel (a) shows the natural line shape at zero velocity, $\chi(\Delta,v = 0)$. In panel (b), effects caused by density ($N$) dependent self (collisional) broadening are added to the widths of each transition. This results in a large Lorentzian broadening of the line. Panel (c) shows velocity-dependent effects on the line shape. The Doppler-broadened line shape $\chi(\Delta,v_\mathrm{z})$ is convolved with a bimodal velocity distribution $g(v)$ that represents the stronger contribution of `slow' atoms to the line shape. The effect is a small additional width. In panel (d) the AS interaction is accounted for by adding a shift modeled using the AS potential $U_\mathrm{vdW}$ for the ground to excited state acting over the entire cell length. The change on the line shape is dramatic for short cell lengths, shifting the peak and causing a long red tail.}
\end{figure*}
\begin{figure}
\includegraphics[scale = 0.7]{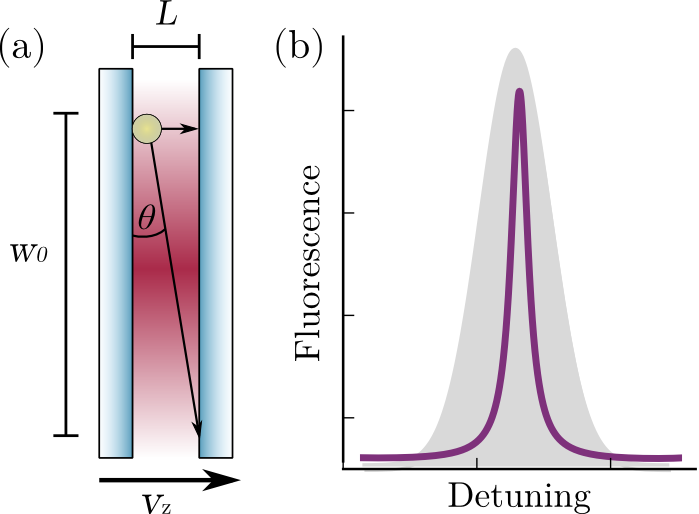}
\caption{\label{fig:dicke_dopp_png}(a) Atomic trajectories in the nanocell that cause Dicke narrowing.  The cell is shorter than the mean free path of the atoms, meaning that atoms traveling perpendicular to the cell walls, with a high velocity in the $z$ direction ($v_\mathrm{z}$) do not have enough time to contribute to the signal, leaving atoms traveling at a small angle to the wall ($\theta$) with a much smaller $v_\mathrm{z}$ as the main component of transmission and absorption spectra. (b) A comparison of Dicke-narrowed absorption spectra (purple line) to the Doppler-broadened spectra (gray area) expected in a 7 cm cell, highlighting the dramatic narrowing caused by velocity-selective effects inside the cell.}
\end{figure}
\begin{figure}
\includegraphics[scale = 0.53]{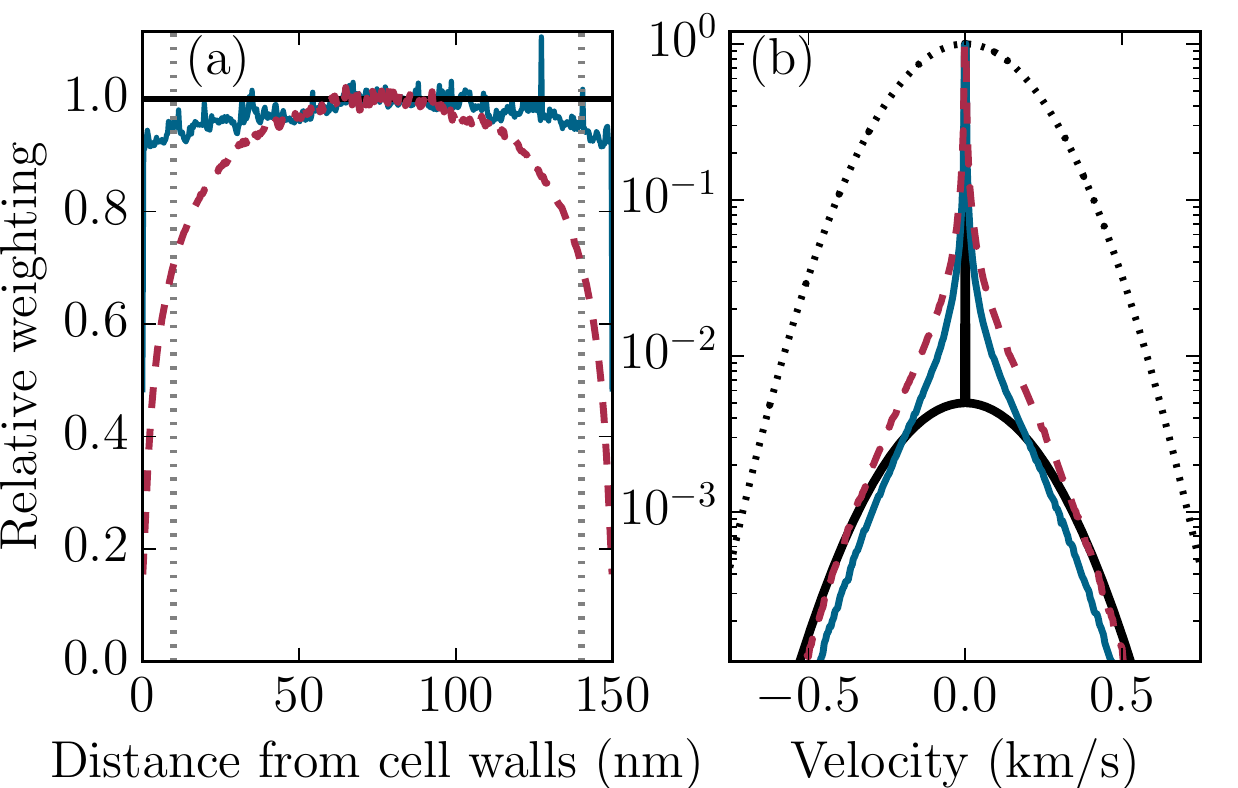}
\caption{\label{fig:MC_sim}Monte-Carlo simulations of atomic trajectories to test the appropriateness of the Dicke narrowing model used in our fitting. Simulations are performed with no AS force present (blue lines) and including the effect of the AS force (dashed red lines). These are compared to the bimodal distribution used in our fitting procedure (thick black lines). In panel (a) the weighting of contributions to the line shape for atomic positions inside the cell is shown. The gray dashed lines mark 10 nm from the cell walls, where the AS force starts to have a considerable effect on the atoms. Panel (b) shows the velocity distribution of atoms. The distribution used in our model (black line) is in reasonable agreement with the simulations for atoms unaffected by the AS force (blue line) and atoms which have been accelerated towards the walls (dashed red line). For comparison, the Maxwell Boltzmann distribution expected in cm scale vapor cells is shown (dashed black line).}
\end{figure}

We analyze the raw data using a model developed to describe the susceptibility of an alkali vapor. The fitting function calculates the detuning-dependent susceptibility $\chi(\Delta)$ of the vapor, then converts it to an absorption spectrum, using a process similar to that outlined in~\cite{ElecSus}. The elements included in the model are outlined in Fig.~\ref{fig:fitting_process}. Each step is illustrated with an example line shape with the most relevant equation displayed beneath it. For comparison, the state of the line shape after (before) each step is shown in red (gray).

The first step is calculating an initial spectrum containing the natural line-width, $\Gamma_0$, for each hyperfine transition, shown in panel (a) on Fig.~\ref{fig:fitting_process}. Each transition is an independent 2 level system. We then sum over all transitions. The single 2-level susceptibility is given by~\cite{bransden2003physics}:
\begin{equation}\label{eqn:2levelchi}
\chi(\Delta) = \frac{\mathrm{i}K}{\Gamma_0/2-\mathrm{i}\Delta},
\end{equation}
where $\Delta$ is the detuning and $K$ is a proportionality constant that includes the dipole matrix element and transition strengths. Atomic data including the natural line-width, dipole matrix element, transition strengths and level splittings are taken from~\cite{ElecSus}.

Panel (b) of Fig.~\ref{fig:fitting_process} shows the effects of collisional broadening, a density-dependent broadening caused by inter-atomic collisions. When an atom in the excited state undergoes a collision with another atom, there is a probability of the excited atom returning to the ground state. This results a reduction of the lifetime leading to a homogeneous broadening of the spectral lines that is linearly dependent on density $\mathcal{N}$~\cite{bransden2003physics}. We model the collisional broadening $\Gamma_\mathrm{col}$ as $\Gamma_\mathrm{col} = \beta\mathcal{N}$, where $\beta$ is a proportionality constant, using values from~\cite{Weller2011a}. The density is calculated from the measured temperature using the vapor pressure formulas in~\cite{ElecSus}. We assume we are within the collision regime where the binary approximation is valid. This is accounted for in the model by including it in the calculation of the susceptibility for single transitions, replacing $\Gamma_0$ in Eq.~\ref{eqn:2levelchi} with~$\Gamma_\mathrm{tot}$, where $\Gamma_\mathrm{tot} = \Gamma_0 + \Gamma_{\mathrm{col}}$. The line shape is then summed over all transitions to generate a spectrum describing the natural line shape broadened by inter-atomic collisions.

Next, we account for velocity-dependent effects, illustrated in panel (c) of Fig.~\ref{fig:fitting_process}. In longer cells, the unrestricted motion of atoms in all directions leads to a Doppler broadening of the spectral lines. However, the tight confinement of the NC restricts the motion of atoms in the propagation direction, altering velocity-dependent shifts seen in the spectra. Atoms with a low velocity vector in the direction of laser beam propagation have a longer interaction time and therefore dominate the signal.

The velocity selected line shape can be understood by exploring the effect of the cell geometry on the atom-light interaction, illustrated schematically in Fig.~\ref{fig:dicke_dopp}(a). The NC is much shorter than the mean free path of the atoms, hence atoms traveling perpendicular to the windows with a high velocity in the direction of beam propagation, $v_z$, collide with the opposite wall frequently, suppressing their contribution to the signal~\cite{Briaudeau1998}. Atoms traveling at a small angle $\theta$ to the wall with a low $v_z$ have a much longer interaction time, and a much smaller Doppler shift. These `slow' atoms undergo more absorption relative to the `fast' atoms, resulting in enhanced sub-Doppler features on top of a small Doppler-broadened background in NC spectra. 

The extreme narrowing is illustrated in Fig.~\ref{fig:dicke_dopp}(b), where the narrowed line shape (purple line) is compared to the Doppler broadened absorption spectrum expected for longer cell lengths (gray area). At low temperatures ($< 100^\circ$C), this narrowing can be so extreme that all individual hyperfine states can be resolved on the Rb D2 line (see Fig.~2(a) in~\cite{Whittaker2014}).

The model accounts for such velocity effects using a phenomenological bimodal velocity distribution for the density of atoms, first described in~\cite{KeaveneyThesis} It describes the number of atoms with velocity $v$, $N_\mathrm{vc}(v)$ as:
\begin{equation}
N_\mathrm{vc}(v) = C(aG_\mathrm{fast}(v) + (1-a)G_\mathrm{slow}(v)),
\end{equation}
where $C = (au\sqrt{\pi}+(1-a)fu\sqrt{\pi})^{-1}$ is a normalization constant, with $u = \sqrt{2k_{\mathrm{B}}T/M}$, the rms velocity of the atoms, and $a$ is a coefficient describing the strength of the contribution of traditional Maxwellian velocities to the signal, and is one of the floating parameters in the fit. $N_\mathrm{vc}(v)$ is split into 2 components, $G_\mathrm{fast}(v)$- the usual Maxwell distribution for describing the Doppler broadened profile, weighted by fitting coefficient $a$:
\begin{equation}
G_\mathrm{fast}(v) = \mathrm{exp}(-v^2/u^2).
\end{equation}
$G_\mathrm{slow}(v)$ describes a narrowed Gaussian profile, representing the atoms traveling at small angles to the walls, that make the larger contribution to the signal:
\begin{equation}
G_\mathrm{slow}(v) = \exp(-v^2/(fu)^2),
\end{equation}
where $f$ is a narrowing factor for the slow atoms. The full distribution is plotted as a black line on Fig.~\ref{fig:MC_sim}(b). The distribution is then convolved with the collisional-broadened line shape.

To test if this bimodal distribution is appropriate, Monte-Carlo (MC) simulation of atomic trajectories was performed and the modified position and velocity distributions were included in the line shape model, as well as a $1/r^3$ atom surface potential that induces a force $F = -\mathrm{d}U_\mathrm{vdW}/\mathrm{d}z$ across the cell. In this simulation we assume that the atom-light interaction region is a box with dimensions $w \times w \times L$, where $w=40~\mu$m is the $1/{\mathrm{e}}^2$ diameter of the excitation laser beam used in the experiment and $L$ is the length of the nanocell. The atoms are randomly placed uniformly across a region which is 4 times larger than the interaction region (4$w~\times~4w~\times L$), to account for atoms transiting into the interaction region during the simulation. The atoms are given random velocities according to Maxwell-Boltzmann statistics, and their motion is simulated over a time (1~$\mu$s) which is large compared to the lifetime of the excited state. We do not include any other decay mechanisms. We neglect desorption events, so atoms that hit the walls or move out of the interaction region are lost. This is justified since desorption processes are expected to happen on time scales much longer than the interaction time~\cite{desorption}.

The results of the simulation are shown in Fig.~\ref{fig:MC_sim}, where dashed red (blue) lines show the results of the model without (with) AS forces present. Panel (a) shows the position-dependent interaction times of atoms inside the cell. Panel (b) compares the MC simulated (blue line) velocity distribution of the atoms to the bimodal distribution used in the model (solid black line) and the Gaussian distribution expected for a Doppler-broadened medium (dotted black line). The bimodal distribution is a reasonable approximation. 

In panel (d) of Fig.~\ref{fig:fitting_process}, we account for the AS interaction by calculating the position-dependent AS shift $\Delta_\mathrm{AS}(r)$ across the entire cell length $L$:
\begin{equation}\label{eqn:AS_pot}
\Delta_\mathrm{AS} =-\left[ \frac{C_\alpha}{r^\alpha}+\frac{C_\alpha}{(L-r)^\alpha }\right],
\end{equation}
where $C_\alpha$ is a coupling coefficient describing the strength of the AS interaction, and $\alpha$ is an exponent describing the power law of the AS interaction. The two terms in Eq.~(\ref{eqn:AS_pot}) simply sum the AS interaction from both cell walls. Both $C_\alpha$ and $\alpha$ are used as floating fit parameters to determine the form of the AS power law. $\Delta_\mathrm{AS}$ is then weighted by the position-dependent interaction times, shown in blue in Fig.~\ref{fig:MC_sim}a. The result is an asymmetric line shape representing shifts across the entire cell weighted by interaction times. This line shape is then convolved with the Dicke narrowed line shape to give the final result of the fitting function. Note that the AS forces slightly modify the position and velocity distribution as shown by the MC simulation results in Fig.~\ref{fig:MC_sim} (red dashed lines), but this is only significant for distances less than 10 nm.

Additional broadening effects that are not currently modeled are accounted for by adding an excess broadening parameter $\Gamma_\mathrm{ex}$ into the fit. We suspect these additional broadenings may be caused by radiation trapping within the nanocell, a future avenue of investigation.

To summarize; the model takes the natural line shape, includes the well-known self broadening, and a fitted additional broadening, $\Gamma_\mathrm{ex}$. We then convolve the line shape with a phenomenological bimodal velocity distribution to account for Dicke narrowing, with a fitted parameter $a$ representing the strength of the contribution of traditional Maxwellian velocity classes. The last step is to convolve with a line shape generated using the AS interaction, fitting for the AS coefficient for the excited state transition $C_\alpha^{5S_{1/2}\rightarrow5P_{3/2}}$ and the exponent of the interaction $\alpha$.

\section{\label{sec:res_diss}Results and discussion}

\subsection{Rubidium D2 line}

\begin{figure}
\includegraphics[scale = 0.56]{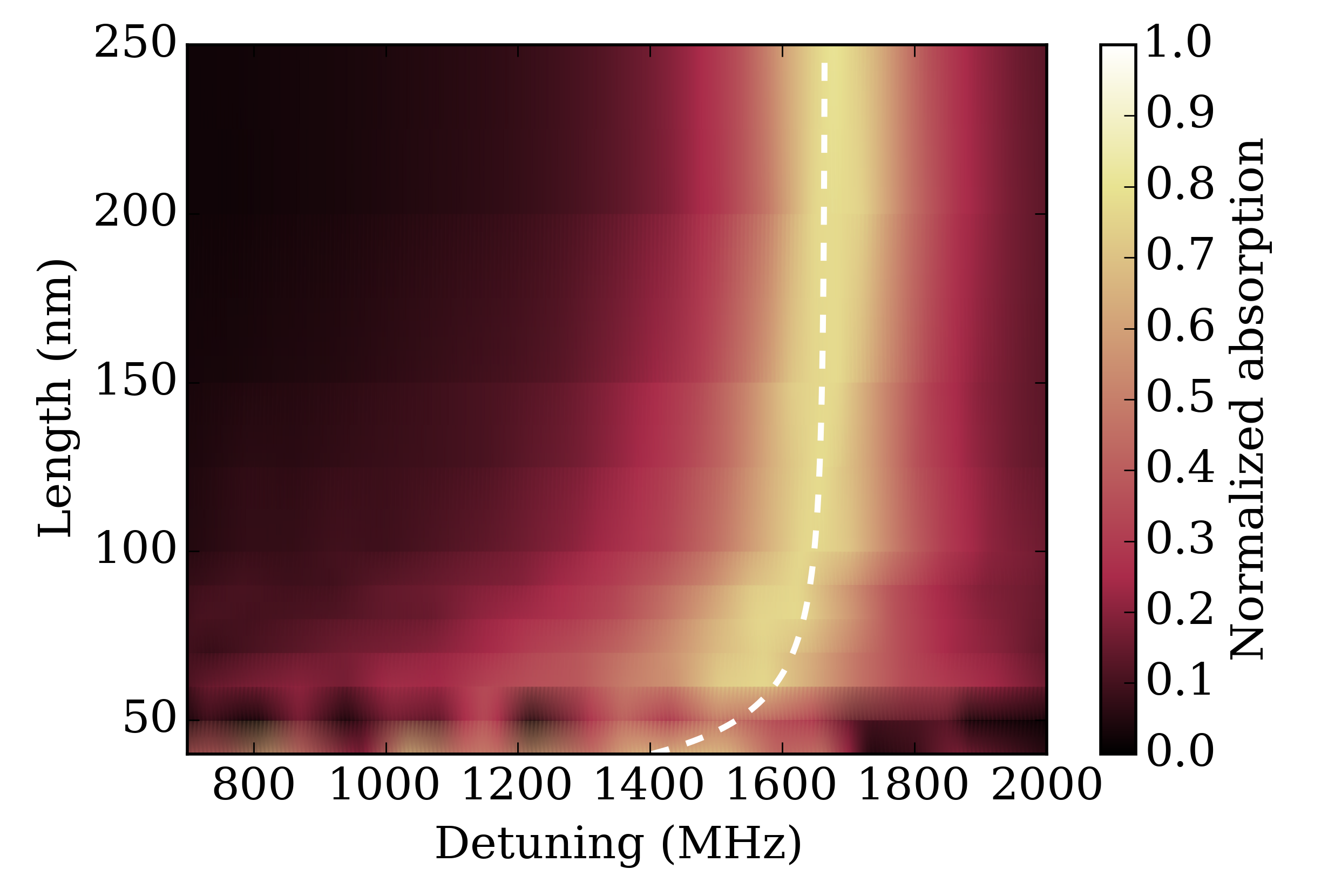}
\caption{\label{fig:AS_cmap}An accumulation of many spectra for different cell lengths. Here we highlight the length dependence of the absorption spectra on the $^{85}$Rb 5$S_{1/2} F_g$ = 2 line to the $^{85}$Rb 5$P_{3/2}$ $F_e$ = 1,2,3 excited state for T = 150$^\circ$C. The expected AS shift for $r_\mathrm{max} = L/2$ is plotted as a white dashed line, using $\Delta_\mathrm{AS} = -C_\alpha^{5S_{1/2}\rightarrow5P_{3/2}}/r_\mathrm{max}^3$. This data set clearly demonstrates the length dependence of the red tail caused by the AS interaction. However in Rb, the red tail and shifts are not appreciable until cell lengths shorter than 80 nm, i.e. maximum AS separations of 40 nm. Hence, fits performed will not be sensitive to the AS interaction outside of this length range.}
\end{figure}

\begin{figure}
\includegraphics[scale = 0.58]{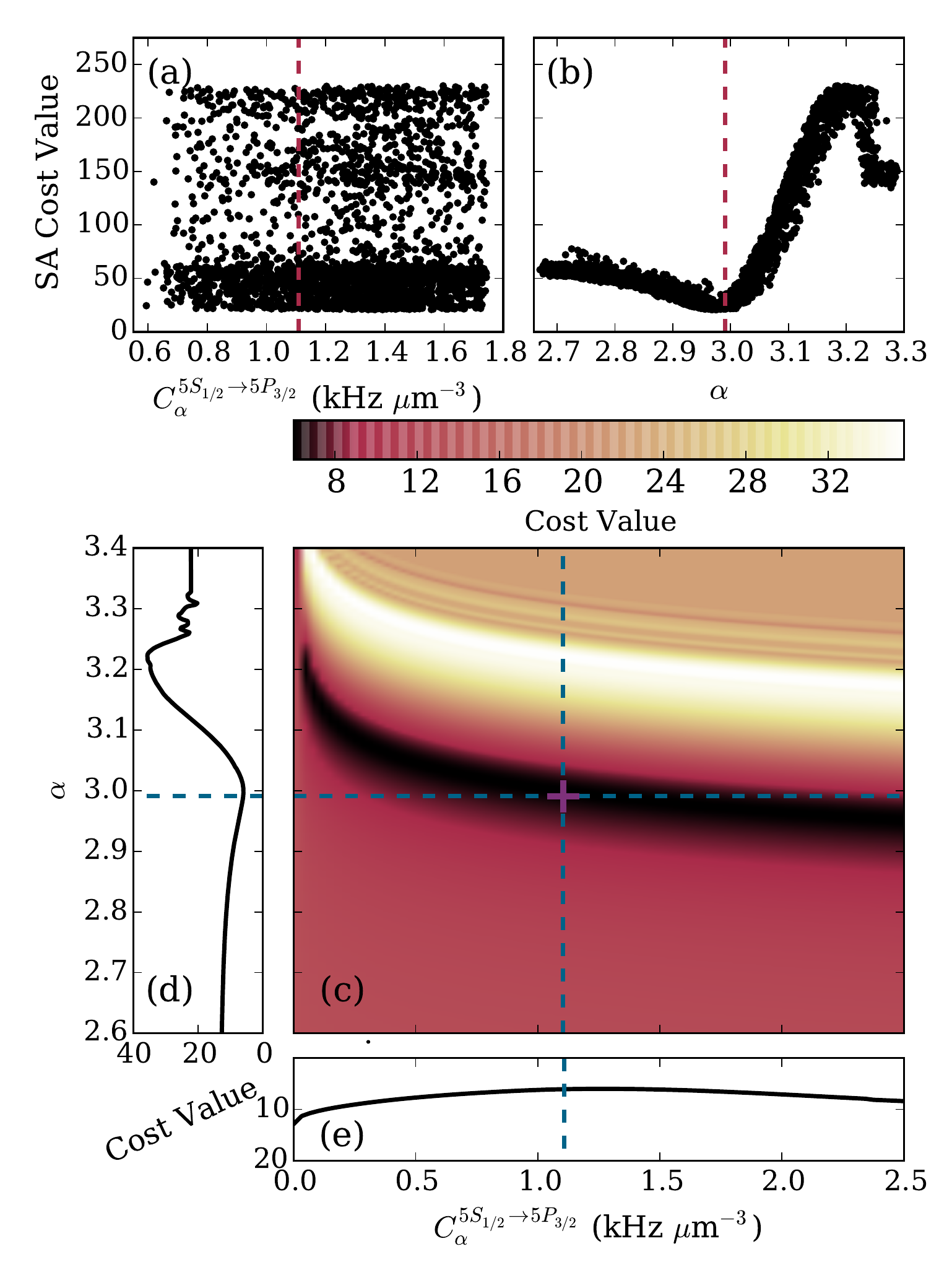}
\caption{\label{fig:rn_comp} Analysis of simulated annealing (SA) fits for a cell of length 80~nm, temperature 125$^\circ$C. (a) Cost values of attempted $C_\alpha$ coefficients tried by the simulated annealing fitting algorithm, with the final fitted value plotted as a red dashed line. There is no clear best value of $C_\alpha$ over the range explored. (b) Cost values of exponent $\alpha$ explored by the SA fitting algorithm. Here, there is a clear region between $\alpha =$ 2.9 and 3.1 where the fit is best. The fitted value found by the SA fitting algorithm is shown as a red dashed line. (c) Parameter space showing cost value across an array of $C_\alpha$ and $\alpha$ with the rest of the fit parameters kept constant. Panels (d) and (e) show the cost values along the fit parameters found by the SA fitting algorithm, highlighting that the region where $\alpha$ is the best fit parameter is very narrow, whereas there is smaller relative variation in the cost value for $C_\alpha$.}
\end{figure}
\begin{figure}
\includegraphics[scale = 0.65]{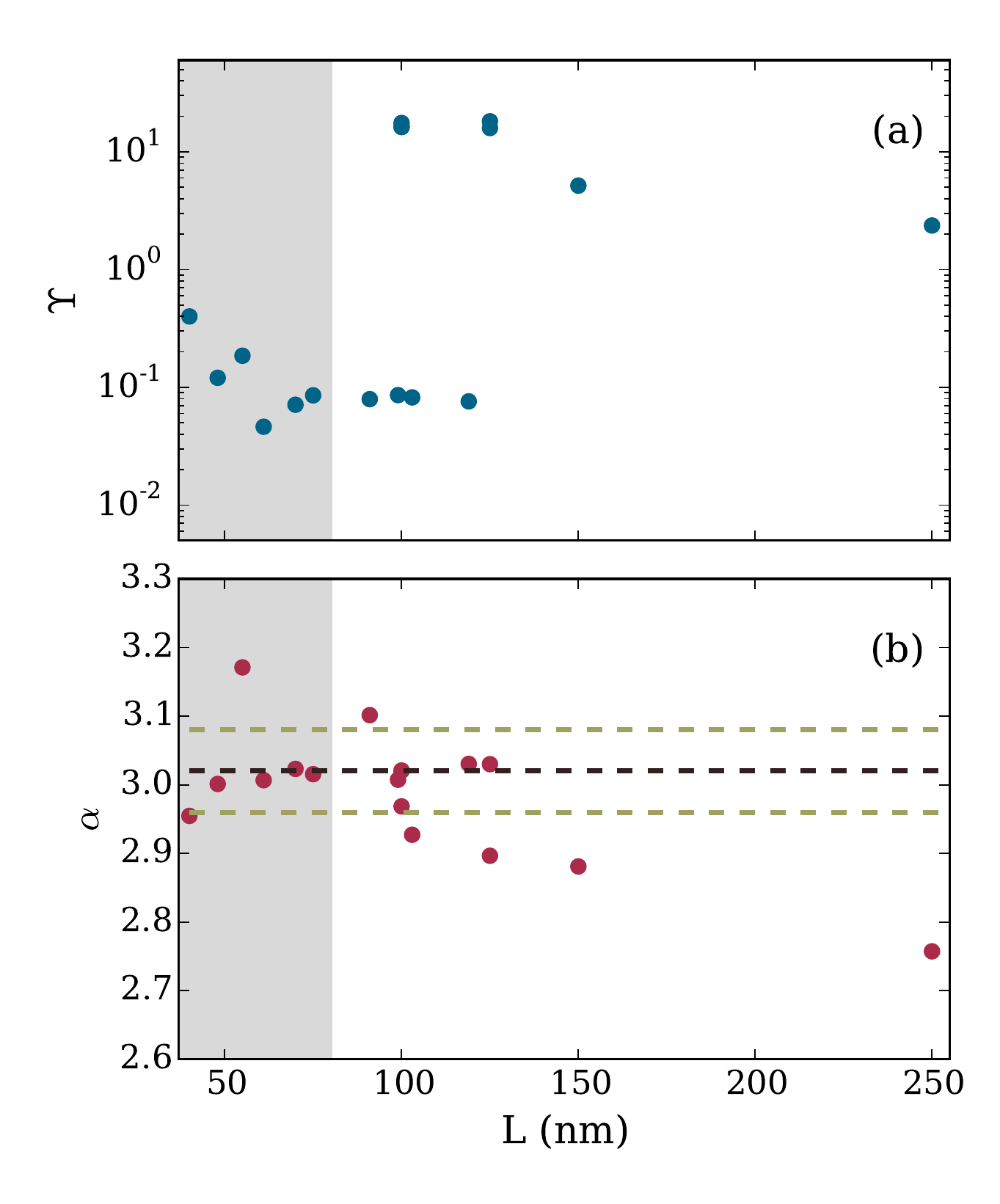}
\caption{\label{fig:as_blobs}Length dependence of goodness of fit parameter $\Upsilon$~\cite{hughes2010measurements}, purple points in panel (a), and the fitted AS exponent $\alpha$, red points in panel (b). The results from longer cell lengths generally have poorer fits than shorter cell lengths where the AS interaction dominates ($L < 80$~nm), highlighted as a gray area. The final value of $\alpha$ = 3.02 $\pm$ 0.06 is plotted as a dashed black line, with the error bars plotted as beige dashed lines.}
\end{figure}

Unlike other experiments that map out the AS interaction using peak shift measurements, we choose to investigate the AS interaction by fitting a full line shape which is most sensitive to smaller AS distances. Spectra are taken over a length range of $L = 40 - 250$~nm at temperatures $T$ = 125, 150, 175 and 200 $^\circ$C. The relative ease of acquisition of spectra also allows us to take many readings over long integration times, increasing the precision of the experiment.

A color map showing the length dependence of the AS induced shifts is shown in Fig.~\ref{fig:AS_cmap}. Data are on the $^{85}$Rb 5$S_{1/2} F_g$ = 2 line to the $^{85}$Rb 5$P_{3/2}$ $F_e$ = 1,2,3 line of the Rb D2 resonance taken at $T$ = 150$^\circ$C. Each horizontal slice shows an experimental spectrum like those in Figs. \ref{fig:raw_rb} (c) and (d), i.e. with background counts subtracted, then normalized such that 1 on the color scale is the peak of the $^{85}$Rb 5$S_{1/2} F_g$ = 2 to the $^{85}$Rb 5$P_{3/2} F_e$ = 1, 2, 3 excited state transitions. The figure demonstrates a clear length dependence of the asymmetric red tail, as expected for the AS interaction. The expected AS shift for atoms in the center of the cell is plotted as a white dashed line. The actual peak shift does not follow exactly the expected $1/r^3$ form because the asymmetric red tail skews the peak location. Hence, line-shape fitting should be more effective than peak detection for determining the functional form of the AS potential.

At long lengths ($L > \lambda/4$), shifts due to the AS interaction on the ground state transitions are very small - on the order of a few MHz. The impact on the line shape is minimal until shorter cell lengths. For the Rb D2 line the asymmetric tail is not appreciable until cell lengths less than 80~nm as evidenced in Fig.~\ref{fig:AS_cmap}. The asymmetric red shift characteristic of the AS interaction is not fully evident until the cell length is below 80 nm with maximum AS separation $r_\mathrm{max}=$ 40~nm. This means our fits are most effective in the length region $r_\mathrm{max} =$ 10~-~40~nm, where there is a sufficiently large red tail to fit to.

After background-count subtraction and normalization, spectra are fitted using the model outlined in the previous section. Examples of fit results on the Rb D2 line for a temperature T = 125$^\circ$C are shown in Figs. \ref{fig:raw_rb} (c) and (d). They show experimental spectra (black points) processed from the raw data shown in Figs. \ref{fig:raw_rb}(a) and (b), fits performed for cell lengths of 60~nm (red line) and 250~nm (blue line).

Upon visual inspection, both spectra have excellent fits. The fit for $L = $ 250~nm in panel (b) has an $\Upsilon$ of 1.7 indicating an excellent fit, although with no AS induced shifts apparent because of the long cell length. The fit for $L = $ 60 nm in panel (a) with an $\Upsilon$ of 0.1 indicates a possible overestimation of the errors involved in the fitting procedure~\cite{hughes2010measurements}. However, as discussed earlier, this is caused by the large background counts for such a short cell length.

To extract an exponent $\alpha$, fit results from all spectra are taken, and an average weighted by the inverse of the $\Upsilon$ of the fit is calculated. The same treatment performed on fitted values of $C_\alpha^{5S_{1/2}\rightarrow5P_{3/2}}$ to determine the coupling coefficient. Ultimately we have determined the best values of $\alpha$ and $C_\alpha^{5S_{1/2}\rightarrow5P_{3/2}}$ to use in future fitting of absorption spectra. Using this method, we extracted a value of $\alpha = 3.02 \pm 0.06$, a value in good agreement with the theoretical expectation of $\alpha  = 3.16$ for the ground state interaction, discussed in section \ref{sec:theory_fit}. We also find a value of $C_\alpha^{5S_{1/2}\rightarrow5P_{3/2}} = 1.4 \pm 0.3$ kHz $\mu$m$^3$, not in agreement with theoretical calculations of $C_\alpha^{5S_{1/2}\rightarrow5P_{3/2}} = 2.1$ kHz $\mu$m$^3$~\cite{priv_comm}. Deviation of the fitted coefficient from theoretical values will be discussed in the next section.

To test the robustness of our fits, we refitted spectra from our data set using a simulated annealing (SA) fitting algorithm~\cite{Kirkpatrick83_SA} instead of the Marquardt-Levenberg (ML) method. The SA fitting algorithm was adapted from the code in~\cite{ElecSus} to use our model of the absorption spectra from section \ref{sec:theory_fit}. To identify the best fit the SA algorithm tries fit parameters and calculates a cost value, calculated as the absolute value of the sum of the theoretical fit subtracted from the dataset. The parameters are varied such that a global optimal fit is found as opposed to a local optimal fit which can returned by a ML fit. The SA algorithm was further adapted to return attempted fit parameters along with the cost value returned for that attempt. 

Fig.~\ref{fig:rn_comp} shows the resulting cost values for $C_\alpha^{5S_{1/2}\rightarrow5P_{3/2}}$, panel~(a), and $\alpha$, panel~(b). Final fitted values from the SA fit, $C_\alpha^{5S_{1/2}\rightarrow5P_{3/2}} = 1.1$ kHz $\mu$m$^3$ and $\alpha = 2.99$ are plotted as dashed red lines on their respective plots. Panel~(a) shows that there is very little variation in the lowest cost value for $C_\alpha^{5S_{1/2}\rightarrow5P_{3/2}}$ over the range of values explored by the SA fitting algorithm, indicating a larger error bar. Conversely, there is a clear region of lowest cost for attempted values of $\alpha$, centred around the fitted value of $\alpha = 2.99$, in agreement with previous analysis.

To explore the possible best fit values, we generated a theoretical parameter space. We kept all parameters aside from $\alpha$ and $C_\alpha^{5S_{1/2}\rightarrow5P_{3/2}}$ constant, then calculated the cost values across an array of $C_\alpha^{5S_{1/2}\rightarrow5P_{3/2}}$ and $\alpha$. The resulting parameter space is shown in Fig.~\ref{fig:rn_comp}~(c). We also show the variation in the cost value along the axes defined by the fitted SA parameters (purple cross), for $\alpha$ in Fig.~\ref{fig:rn_comp}~(d) and for $C_\alpha^{5S_{1/2}\rightarrow5P_{3/2}}$ in Fig.~\ref{fig:rn_comp}~(e). The selections are highlighted on panel (c) as blue dashed lines. Again, the region of lowest cost value for $\alpha$ is much narrower than for $C_\alpha^{5S_{1/2}\rightarrow5P_{3/2}}$, showing that this method of fitting spectra is more effective in identifying the exponent, $\alpha$, of the AS interaction. However, the larger uncertainty in best cost value for $C_\alpha^{5S_{1/2}\rightarrow5P_{3/2}}$ in the parameter space shows that this method is less suited to find an accurate value of $C_\alpha^{5S_{1/2}\rightarrow5P_{3/2}}$.

The length dependence of the $\Upsilon$ (blue points) of each ML fit with the corresponding fitted values of $\alpha$ (red points) are plotted in Figs. \ref{fig:as_blobs} (a) and (b) respectively. In panel (b), the experimentally determined value is highlighted with a black dashed line, with the standard deviation plotted as gray dashed lines. Comparing with Fig.~\ref{fig:rn_comp}~(c), where there is a range of $\alpha$ where the cost value is minimal, between $\alpha = $~3.2 and~2.9, we can see that the fitted $\alpha$ from the ML fitting method all lie within this length range, showing that the ML fit explored the whole region of lowest cost values of $\alpha$, and still returns final fit parameters clustered around our final quoted value of $\alpha$. In panel (a) the $\Upsilon$ of the fit is generally larger for longer cell lengths, where the AS shift is not appreciable. Fitted values of $\alpha$ and $C_3^{5S_{1/2}\rightarrow5P_{3/2}}$ calculated at these longer lengths have very little bearing on the nature of the AS interaction, and a weighted average using $\Upsilon$ will decrease their impact on the final measured value. For shorter cell lengths $L < 80$~nm, the AS interactions have much stronger effects on the line shape also coinciding with smaller values of $\Upsilon$, with most fitted values of $\alpha$ in good agreement with the final average value and its associated errors.

\subsection{Cesium D1 line}
\begin{figure}
\includegraphics[scale = 0.5]{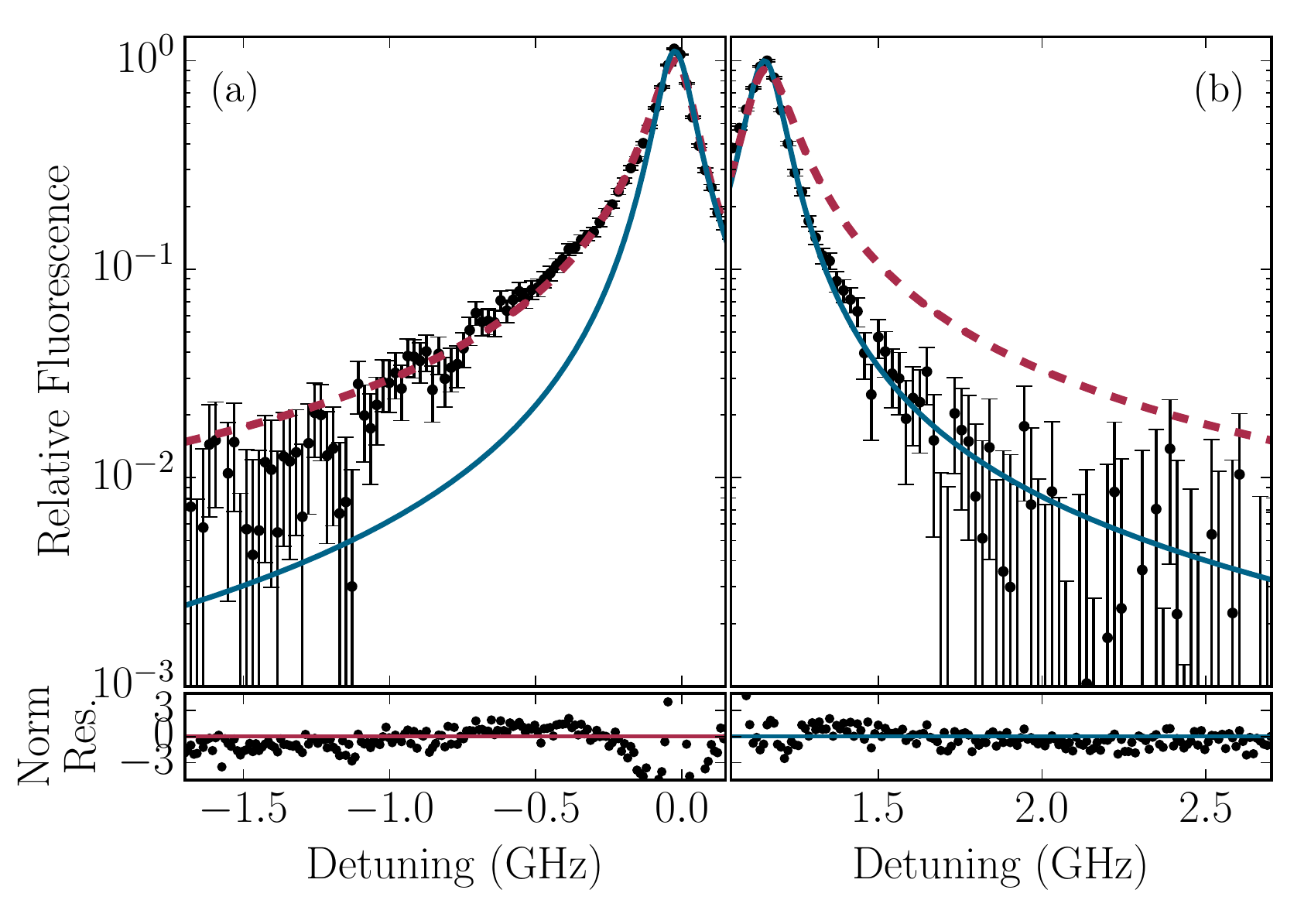}
\caption{\label{fig:red_blue_comp}Comparison of the red (a) and blue (b) wings of the  Cs 6$S_{\frac{1}{2}} F_g$ = 3 to 6$P_{\frac{1}{2}} F_e$ = 4 transition. Zero detuning is the resonance frequency of the line. There is a remarkable difference between the two wings, showing what a profound impact the AS interaction has on the line shape. The data (black points) are compared to the red wing as fitted by our model (dashed red line, mirrored in panel (b) for comparative purposes) and a Lorentzian fit to the blue wing (blue line). A Lorentzian fit is sufficient to model the blue wing, because Doppler broadening is suppressed by Dicke narrowing of the resonance. 
}
\end{figure}

To corroborate the results found in Rb, and to test the model with other alkali-metal atoms, the experiment was performed in a Cs NC on the Cs D1 894~nm line, where large hyperfine splittings (9 GHz for the Cs ground states and 1.2 GHz on the excited states). Each hyperfine transition is separated such that it can be considered in isolation from all others, even in the presence of a large AS interaction. This allows the reliable measurement of shifts with a precision of a few MHz.

Whilst the ability to address individual transitions gives exceptionally high resolution, the technical challenges of the Cs D1 line are greater than the Rb D2 line, primarily due to the transition wavelength. Firstly, the quantum efficiency of Si, used for detection in the SPCMs, is much lower at 894~nm than at 780~nm and hence the detectors are not as sensitive. Secondly, high transmission, narrow bandpass filters are currently not available at 894~nm meaning relatively more thermal background photons (of which there are more in absolute number compared to 780~nm) reach the detector. As a result of these two factors, we limit our investigation to nanocell thicknesses greater than 100~nm for the Cs D1 line in order to generate spectra with sufficiently high signal-to-background ratio.

Using the same methods as those to extract parameters for Rb, we find a value of $C_3^{6S_{1/2}\rightarrow6P_{1/2}} = (1.9 \pm 0.1)$ kHz $\mu$m$^3$, a value that, similar to the Rb coefficient, is in disagreement with theoretical expectations of $C_\alpha^{6S_{1/2}\rightarrow6P_{1/2}} = 1$ kHz $\mu$m$^3$~\cite{chevrollier1992high}.

In Fig.~\ref{fig:red_blue_comp}, we present a comparison of the red and blue wings of AS shifted transitions, the Cs 6$S_{1/2} F_g$ = 3 to 6$P_{1/2} F_e$ = 3, panel (a), and $F_e$ = 4, panel (b). The blue line on both plots is a Lorentzian fit, i.e. one that excludes the AS interaction. The red line on panel (a) is a fit including the AS interaction, and is mirrored in panel (b), highlighting the asymmetry of the AS interaction. These spectra show that broadening effects are exclusively on the red wing.

\section{Conclusion}

We measure and analyze the resonance line shapes of atoms confined between sapphire plates with separations between 30 and 250~nm. The lineshapes are asymmetric due to the AS interaction. 

We fit the absorption line shapes in this length range, with an AS interaction of the general form $-C_\alpha/r^\alpha$. We found the best fitted values by fitting absorption spectra taken at multiple temperatures over the aforementioned length ranges, a method that greatly enhances the precision of our experiment. The resulting fitted parameters were averaged with a weighting according to the reduced chi squared (denoted as $\Upsilon$) of the fit. We find an averaged value of $\alpha$ = 3.02 $\pm$ 0.06, a value in good agreement with theoretical expectation for the atom surface interaction inside the near field, and confirmed by other experiments \cite{Laliotis15}. 

We also found a fitted value of $C_\alpha^{5S_{1/2}\rightarrow5P_{3/2}} = 1.4\pm 0.1$~kHz~$\mu$m$^3$, a value not in agreement with theoretical expectations of 2.1~kHz~$\mu$m$^3$. We postulate that the discrepancy between experiment and theory may be caused by rubidium adsorbed on the interior surface of the cell, interfering with the surface permittivity and therefore the reflection coefficient used to calculate  $C_3^{5S_{1/2}\rightarrow5P_{3/2}}$~\cite{Sukenik1993patch_CP,Eizner_CPvdW}. The effect of metallic layers on the surface affects only the $C_3^{5S_{1/2}\rightarrow5P_{3/2}}$ coefficient, leaving the exponent of $r^\alpha$ unchanged~\cite{Eizner_CPvdW}. This leads us to conclude that vapour cells may not be an appropriate medium for the accurate measurement of $C_3$ without rigorous theoretical treatment to account for the effects of surface atoms, currently beyond the scope of this experiment. There may be more complex interactions occurring inside the cell~\cite{bloch_comment}, such as additional motional effects between photon absorption and fluorescence, multiple reflections and surface charges. However, we have demonstrated that a simple model is sufficient to describe the power law of $1/r^3$ for spectra over the length scales investigated. 

In future we intend to utilize the methods developed herein to aid future investigations on the AS interaction, including the search for bound states close to the surface~\cite{Lima2000} and investigate higher-excited D states~\cite{Failache1999}. To further characterize our system, we also hope to look at the effects of radiation trapping~\cite{Fioretti1998415}. Finally, we will use the model developed to aid in the fabrication and testing of nanocells being developed in Durham.

\section{Acknowledgments}

The authors would like to thank D. Bloch for stimulating our interest in some of the topics presented here and K. N. Jarvis for assistance in data acquisition. We acknowledge financial support from Durham University and the ESPRC (Grant EP/L023024/1). The data presented in this paper are available \href{http://dx.doi.org/10.15128/s1784m356 }{online}.

\bibliography{followup_bib16}
\bibliographystyle{apsrev}

\end{document}